\documentclass[fleqn,twoside]{article}
\usepackage{espcrc2}
\usepackage{epsfig}

\usepackage{graphicx}
\def\dj{d\hskip-.4em\hbox{\char'26}\hskip-.1em}
\def\Dj{D\hskip-.8em\lower.4ex\hbox{\char'26}\hskip.3em}

\newcommand{\AmS}{{\protect\the\textfont2
  A\kern-.1667em\lower.5ex\hbox{M}\kern-.125emS}}

\title{Large-Scale Production of Monitored Drift Tube Chambers for the
ATLAS Muon Spectrometer}

\author{
F.~Bauer\address[ad1]{Max-Planck-Institut f{\"u}r Physik, F{\"o}hringer Ring 6,
D-80805 Munich, Germany}\thanks{Permanent address: CEA Saclay, F-91911
Gif-sur-Yvette, France.}, S.~Horvat\addressmark[ad1]\thanks{Permanent
address: Institut Ru\dj er 
Bo\v skovi\' c, 10001 Zagreb, Croatia.}, O.~Kortner\addressmark[ad1],
H.~Kroha\addressmark[ad1], 
A.~Manz\addressmark[ad1], S.~Mohrdieck\addressmark[ad1], 
R.~Richter\addressmark[ad1],
V.~Zhuravlov\addressmark[ad1]\thanks{Now at CERN, CH-1211 Geneva 23, Switzerland and
JINR, Dubna, 141980 Moscow Region, Russia.}}

\begin{document}

\begin{abstract}
\noindent
Precision drift tube chambers with a sense wire positioning accuracy of 
better than $20~\mu$m are under construction for the ATLAS
muon spectrometer. $70\%$ of the 88 large chambers for the outermost
layer of the central part of the spectrometer have been assembled.
Measurements during chamber construction of the positions of the sense wires and
of the sensors for the optical alignment monitoring system 
demonstrate that the requirements for the mechanical precision of the chambers 
are fulfilled.
\vspace{1pc}
\end{abstract}

\maketitle

\section{Introduction}
\label{intro}
\noindent
The muon spectrometer~\cite{muon} 
is one of the main characteristics of
the ATLAS experiment at the Large Hadron Collider (LHC).
It has been designed to measure muon momenta in the range from 10 to 1000~GeV 
with an accuracy of 3 to $10\%$ over a pseudo-rapidity range of $|\eta |\le 2.7$.
The muon trajectories in the 0.5~T toroidal magnetic field of superconducting
air-core magnets are measured by three stations of precision
drift chambers, the Monitored Drift Tube (MDT) chambers. The
MDT chambers consist of a pair of triple layers of drift tubes (quadruple layers
in the innermost station) mounted on a light aluminum space frame. 
The aluminum drift tubes of 30~mm diameter 
contain $50~\mu$m diameter gold-plated W-Re sense wires and
are operated with Ar:CO$_2$~(93:7) gas mixture at a pressure of 3 bar
and a gas gain of $2\cdot 10^4$.
\begin{figure}[b!]
%\unitlength1cm
%\begin{picture}(6,4.5)
%\put(0.7,-0.8){\includegraphics[width=6cm]{xtomo_bos_2003_bw_3.eps}
%\put(0.7,-0.8)
\vspace{-12mm}

\mbox{\epsfig{file=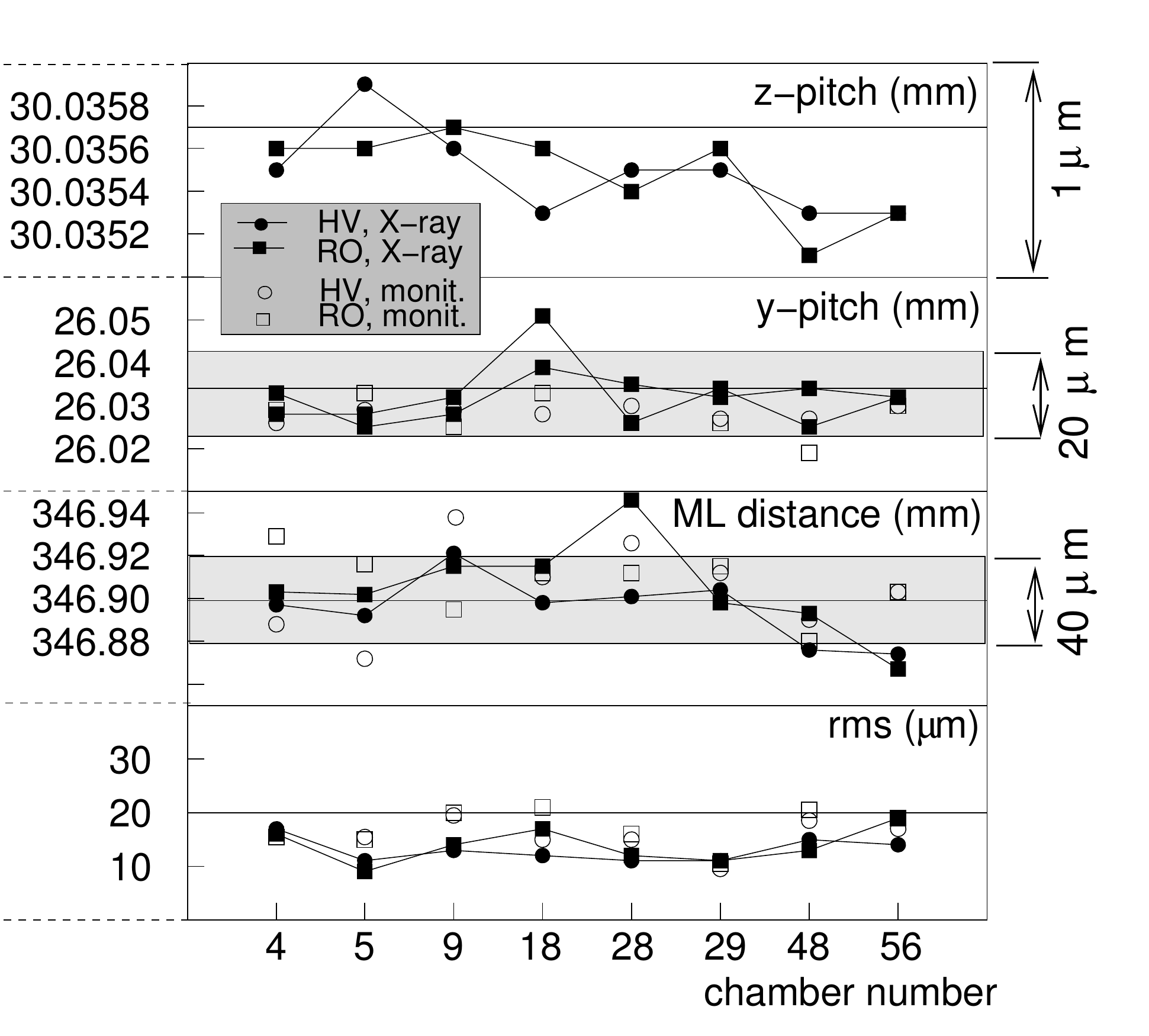,width=1.03\linewidth}}
%\end{picture}
\vspace{-14mm}

\caption{X-ray measurements of the geometrical parameters of the wire grid
at the high-voltage (HV) and the readout (RO) end of the MDT chambers in
comparison with the optical measurements during construction.
The last row shows the rms values of the X-ray measurements of the wire positions
with respect to the measured wire grid (full symbols) and to the wire positions
determined during construction (open symbols). 
The horizontal lines represent the nominal values.}
\label{fig.xtomo}
\end{figure}
\begin{figure}[b!]
%\unitlength1cm
%\begin{picture}(6,4.3)
%\put(1.7,4.6){a)}
%\put(0.7,-0.8){\includegraphics[width=6cm]{rms-wires-bw.eps}}
%\put(0.7,-0.8)
\vspace{-9mm}

\mbox{\epsfig{file=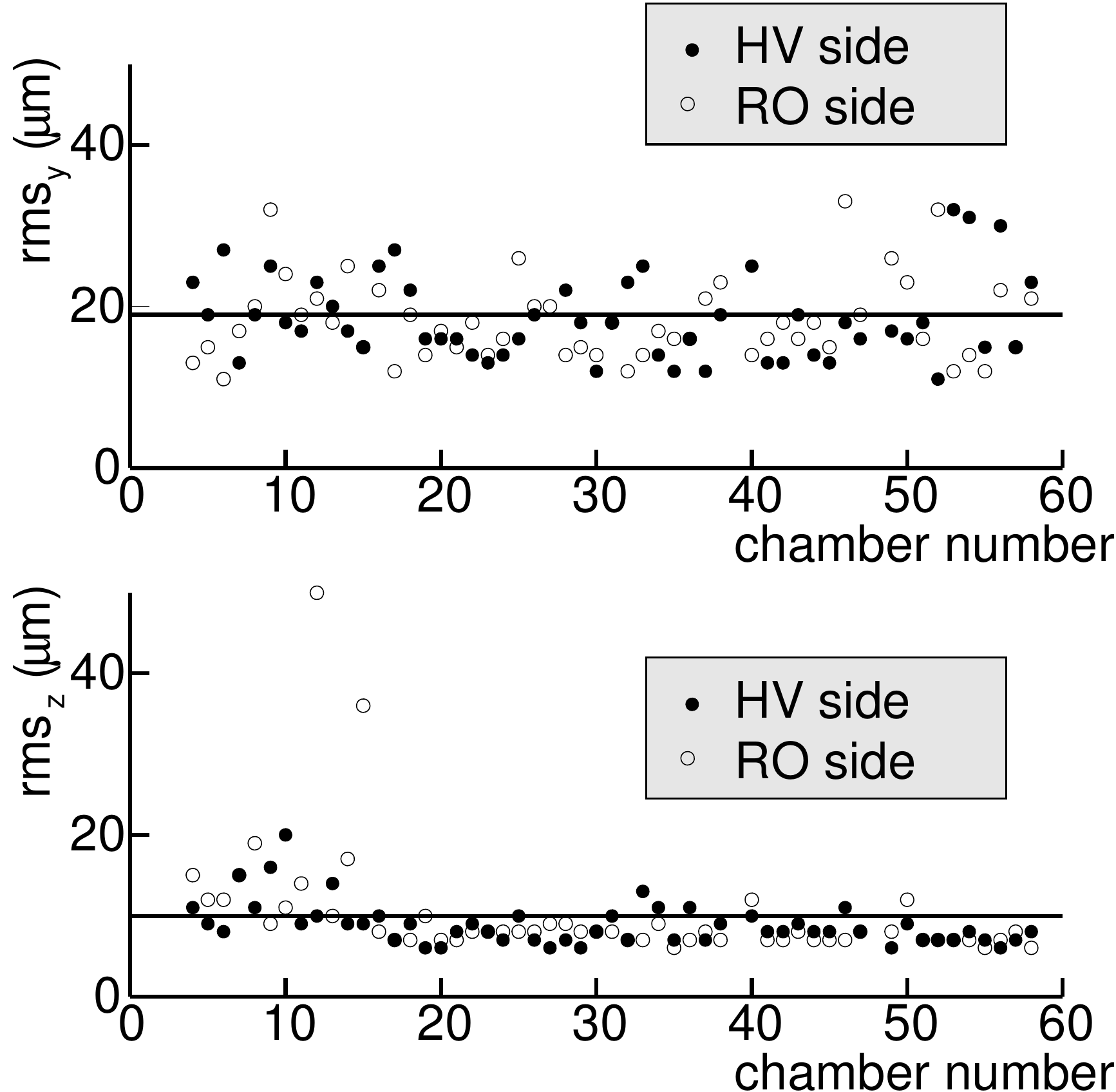,width=\linewidth}}
%\put(1.7,1.7){b)}
%\end{picture}
\vspace{-12mm}

\caption{The rms values of the residuals of the wire positions reconstructed during
chamber construction with respect to the nominal wire grid 
in the vertical ($y$) and the horizontal ($z$) coordinate
at the high-voltage (HV) and the readout (RO) end of the chambers.
The horizontal lines represent the average values.}
\label{fig.wires}
\end{figure}
\begin{figure}[b!]
%\unitlength1cm
%\begin{picture}(6,5.2)
%\put(1.6,5.3){a)}
%\put(0.3,-1.3){\includegraphics[width=7cm]{poster_dposi-bw.eps}}
%\put(0.3,-1.3)
\vspace{-16mm}

\mbox{\hspace{-4mm}\epsfig{file=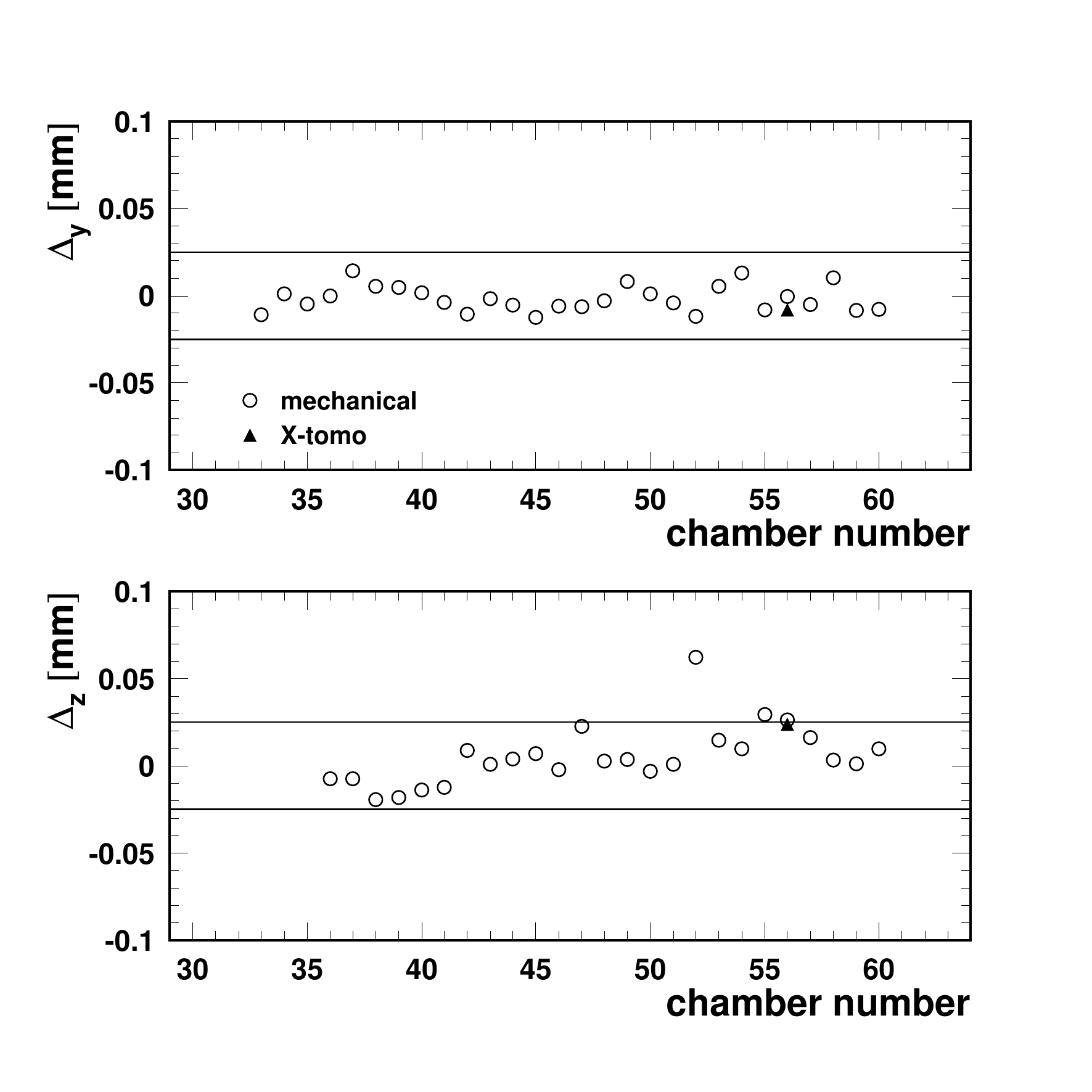,width=8.8cm}}
%\put(1.6,1.8){b)}
%\end{picture}
\vspace{-18mm}

\caption{Deviations of the measured vertical ($y$) and  
horizontal ($z$) alignment platform positions from the
nominal values as a function of the chamber number.
The required tolerances are indicated by horizontal lines.
The X-ray measurements for one chamber are shown as black triangles.}
\label{fig.platform}
\end{figure}

The muon chambers have to provide a position resolution of $40~\mu$m.
With a single-tube resolution of $100~\mu$m~\cite{X5test},
this is achieved by requiring a positioning accuracy of the sense wires
in a chamber of better than $20~\mu$m (rms). The relative positions of the
MDT chambers in the ATLAS detector are continuously measured by an
optical alignment monitoring system~\cite{muon} which is designed to provide 
misalignment corrections to the muon sagittae with an accuracy of $30~\mu$m (rms).
The precision of the misalignment corrections is directly proportional to
the positioning accuracy of the optical sensors mounted on the chambers
which is required to be $\pm 25~\mu$m with respect to the sense wires.

1200 MDT chambers covering an active area of 5000~m$^2$ 
are under construction for the ATLAS muon spectrometer.
The MPI Munich is responsible for the production of 88 chambers
for the outermost layer of the central part of the spectrometer
covering $15\%$ of the active area.
Each chamber contains 432 drift tubes of 3.8~m length. 
$70\%$ of the chambers have already been assembled.

\section{Chamber Assembly and Monitoring}
\label{asse}
The MDT chamber assembly procedure has been described in~\cite{mpinote},\cite{mpipub}.
The required mechanical precision of the chambers is achieved in two steps.
During the assembly of the drift tubes, the sense wires are centered
at the tube ends with an accuracy of $7~\mu$m (rms) as verified by X-ray
measurements of every drift tube~\cite{mpinote}.
In the second step, the 72 drift tubes of a tube layer are
positioned on the chamber assembly table with an accuracy of $5~\mu$m (rms)~\cite{mpipub}
at the tube ends verified by measurements with mechanical feeler gauges~\cite{mpinote}.

The tube layers positioned on the assembly table are consecutively glued to the
space frame which, for this purpose, is positioned on the table with an accuracy 
of $\pm 5~\mu$m with respect to the tubes. The effects of glue shrinkage on the geometry are
anticipated.
The gravitational deformations of the space frame supported on the table 
are measured by optical sensors (RASNIK imaging systems developed by NIKHEF
consisting of an illuminated grid pattern, a focusing lens and a CCD camera)
and are compensated during glueing of the chambers by computer-controlled 
pneumatic actuators.

The wire positioning accuracy of the chambers is determined by the precision
of the assembly tools. For about $15\%$ of the chambers, the assembly precision 
is verified by wire position measurements of $3~\mu$m (rms) precision with an X-ray 
scanning device at CERN~\cite{xtomo}. The measured geometrical parameters of the wire 
grid, the wire pitch in the directions parallel ($z$) and vertical ($y$) 
to the assembly table and the distance between the triple layers (ML),
are reproducible and in good agreement with the nominal values (see Fig.~\ref{fig.xtomo}). 
The wire positioning accuracy fulfills the requirement;
the average accuracy of the chambers X-rayed is $14~\mu$m (see Fig.~\ref{fig.xtomo}). 

In order to monitor the accuracy of the chambers during the construction,
the geometrical chamber parameters are measured at the four
corners of the space frame with dedicated RASNIK sensors which provide a measurement 
accuracy of $10~\mu$m.
The optical measurements of the $y$ pitch and of the triple layer (ML) distance
are in good agreement with the nominal values and with the X-ray measurements 
as shown in Fig.~\ref{fig.xtomo} and are found to be
reproducible during chamber construction within $15~\mu$m and $20~\mu$m (rms), respectively. 

By combining the optical measurements of the layer and triple layer positions
with the additional information from the tube position measurements on the assembly table and the
wire positions in the individual drift tubes, the wire coordinates
within a chamber can be determined. The rms values of the residual distributions
of the reconstructed wire positions with respect to the nominal values and with respect
to the X-ray measurements are shown as a function of the chamber number 
in Figs.~\ref{fig.wires} and \ref{fig.xtomo}, respectively.
The $y$ and $z$ coordinates of the wires are reconstructed with average accuracies
of $18~\mu$m and $14~\mu$m, respectively.
\section{Alignment Sensor Positioning}
\label{plat}
\noindent
The sensors for the optical alignment monitoring system are mounted on
aluminum platforms glued to the triple layers with precise jigs.
The positions of the platforms with respect 
to the sense wires are measured with mechanical feeler gauges
on the assembly table with accuracies of $\pm 5~\mu$m and $\pm 10~\mu$m in $y$ and $z$
direction, respectively. The measured positions are found to be within the required 
tolerances and agree with X-ray measurements of the platform positions at CERN
(see Fig.~\ref{fig.platform}). The angular orientations of the platforms are
measured with the same method with accuracies of $\pm 50-100~\mu$rad, sufficient
to determine them within the required tolerances.
\section{Conclusions}
\label{conc}
The construction of precision drift tube chambers for the ATLAS muon spectrometer
is well advanced. The chambers fulfill the requirements for the positioning
of the sense wires and of the sensors for the optical alignment monitoring system
as verified by measurements during chamber construction.
\section*{Acknowledgements}
\label{ack}
We thank the X-ray tomograph group at CERN for the careful measurement of our
chambers and for providing their results to us.
\noindent

\end{document}